\documentclass[utf8]{frontiersFPHY} 

\usepackage{url,hyperref,microtype,subcaption}
\usepackage[onehalfspacing]{setspace}
\usepackage{braket}

\def\keyFont{\fontsize{8}{11}\helveticabold }
\def\firstAuthorLast{M. Tokieda and K. Hagino} 
\def\Authors{M. Tokieda\,$^{1}$ and K. Hagino\,$^{2,*}$}
\def\Address{
$^{1}$Department of Physics, Tohoku University, Sendai 980-8578, Japan \\
$^{2}$Department of Physics, Kyoto University, Kyoto 606-8502, Japan}
\def\corrAuthor{K. Hagino}
\def\corrEmail{hagino.kouichi.5m@kyoto-u.ac.jp}

\begin{document}
\onecolumn
\firstpage{1}

\title[Time-dependent approaches to open quantum systems]
{Time-dependent approaches to open quantum systems} 

\author[\firstAuthorLast ]{\Authors} 
\address{} 
\correspondance{} 

\extraAuth{}

\maketitle

\begin{abstract}

Couplings of a system to other degrees of freedom (that is, environmental degrees of freedom) 
lead to energy dissipation when the number of environmental 
degrees of freedom is large enough. Here we discuss quantal treatments for such energy dissipation. 
To this end, we discuss two different time-dependent methods. One is to introduce an effective time-dependent 
Hamiltonian, which leads to a classical equation of motion as a relationship among expectation 
values of quantum operators. We apply this method to 
a heavy-ion fusion reaction 
and discuss the role of dissipation on the penetrability of the Coulomb barrier. 
The other method is to start with a Hamiltonian with environmental degrees of freedom and 
derive an equation which the system degree of freedom obeys. 
For this, we present a new efficient method to solve coupled-channels equations, which 
can be easily applied even when the 
dimension of the coupled-channels equations is huge. 

\tiny
 \keyFont{ \section{Keywords:} 
open quantum systems, quantum friction, Caldeira-Leggett model, 
barrier transmission, fusion reactions} 
\end{abstract}

\section{Introduction}

Open quantum systems are ubiquitous in many branches of science. In general, 
a system is never isolated but couples to other degrees of freedom, which 
are often referred to as environment. The couplings to the environmental 
degrees of freedom can strongly affect the dynamics of the system. When the number 
of environmental degrees of freedom is huge, the couplings lead to energy 
dissipation. It has been demonstrated by Caldeira and Leggett that such 
couplings suppress the tunneling rate \cite{CL83}, 
going into a transition from 
quantum to classical regimes. 
In nuclear physics, it has been well-known  
that a large amount of the relative energy and angular momentum are 
dissipated in collisions of heavy nuclei at energies close to the Coulomb 
barrier, known as deep inelastic collisions \cite{FL96}. In this case, 
the dissipation occurs due to the couplings between the relative motion 
of two colliding nuclei and nucleonic degrees of freedom in those nuclei. 
A classical Langevin equation \cite{Abe96} 
has been successfully applied to describe 
such collisions \cite{FL96}. 
The Langevin approach has also been employed in order to discuss fusion 
reactions for syntheses of superheavy elements 
\cite{Abe00,Abe02,Aritomo04,fbd05,ZG15,H18}. 

The classical 
Langevin approach, by definition, is not applicable at energies around 
the Coulomb barrier, at which quantum effects play an important 
role \cite{Takigawa04,Ayik05}. 
One can then ask: what is a quantum model which in the classical 
limit is equivalent to 
a classical Langevin equation? 
There are two approaches to address this question. 
One is to use a phenomenological quantum friction model, 
in which the expectation values of operators obey the 
the classical equation of motion with friction \cite{CK1,CK2,KO72,AL75}. 
Recently, we solved such quantum friction Hamiltonians with a time-dependent 
wave packet approach in order to discuss the effect of friction on 
quantum tunneling \cite{TH17}. 
The other approach to a quantum Langevin equation is to start from 
a system-plus-bath Hamiltonian, that is, 
a Hamiltonian which consists of the system and the environmental degrees 
of freedom, and eliminate the environmental degrees of freedom. 
For instance, one can employ the Caldeira-Leggett Hamiltonian \cite{CL83}, 
since the classical Langevin equation can be 
derived from it \cite{FL96,Abe96}. 
This approach is more microscopic, 
thus a computation would be more involved, than the quantum friction model. 
It has been known that, in the Markovian limit, 
the time evolution for the reduced density matrix for the system degree of freedom 
in general takes the so called Lindblad form \cite{Lindblad76,Pearle12}. 

In this paper, we discuss both of these two approaches for open quantum 
systems, from a point of view of time-dependent method. 
In the next section, we first discuss the phenomenological quantum 
friction models using a time-dependent wave packet approach. 
We apply them to heavy-ion fusion reactions around the 
Coulomb barrier, and discuss a role of friction in fusion dynamics. 
In Sec. III. we solve the Calderira-Leggett Hamiltonian using a 
time-dependent coupled-channels approach. Using a quantum damped harmonic 
oscillator, we discuss how one can deal with a large number of degrees of 
freedom. A summary of the paper is then given in Sec. IV. 

\section{Phenomenological quantum friction models}

We first consider a phenomenological approach to quantum friction.
In this approach, one treats the environmental degrees of freedom implicitly
and introduces a phenomenological Hamiltonian  with which
the classical equation of motion with a frictional force
is reproduced as expectation values.
For this purpose, several model Hamiltonians have been proposed 
so far \cite{CK1,CK2,KO72,AL75}.
Among these, we focus in this paper on the one introduced by Kostin \cite{KO72}. 

Consider a particle of mass $m$ moving in a 
one dimensional space $q$ under the influence of a potential $V(q)$.
With a friction coefficient $\gamma$, the phenomenological
Schr\"odinger equation in the Kostin model is given by \cite{KO72} 
\begin{equation}
i\hbar \frac{\partial}{\partial t} \psi(q,t) = 
\left[ - \frac{\hbar^2}{2m} \frac{\partial^2}{\partial q^2} + V(q) + \gamma S(q,t) \right]
\psi(q,t),
\label{sch_1dkostin0}
\end{equation}
where $S(q,t)$ is the phase of the wave function, 
$\psi(q,t) = |\psi(q,t)| \exp(i S(q,t) / \hbar)$.
From this equation, 
it is easy to confirm that one can derive the equation of motion with a frictional force
\begin{equation}
\frac{d}{dt} \langle p \rangle = - \gamma \langle p \rangle
- \left\langle \frac{dV}{dq} \right\rangle,
\label{eom0}
\end{equation}
as is desired.
Here, the expectation value of an operator $\mathcal{O}$
is denoted as $\langle \mathcal{O} \rangle =
\int dq \,  \psi^*(q,t) \mathcal{O} \psi(q,t)$
and $p$ is the momentum operator.

When one simulates an energy dissipation in nuclear collisions
by means of friction, a frictional force
should be active only when the colliding nuclei are close to each other. 
In other words, one needs to deal with a coordinate dependent
friction coefficient, $\gamma=\gamma(q)$.
An extension of the Kostin model along this line has been proposed in Refs. \cite{IKG75,BM14}, 
with which the modified Schr\"odinger equation is given by
\begin{equation}
i\hbar \frac{\partial}{\partial t} \psi(q,t) =
\left[ - \frac{\hbar^2}{2m} \frac{\partial^2}{\partial q^2} + V(q) + \int^q dq_1 \, \gamma(q_1) \frac{\partial}{\partial q_1}S(q_1,t) \right]
\psi(q,t).
\label{sch_1dkostin}
\end{equation}

To apply the phenomenological model to realistic collision problems,
one further needs an extension to a three dimensional space,
$\vec{q} = \vec{q}(r,\theta,\phi)$.
To this end, we first expand the wave function with
the Legendre polynomials $P_{l}(x)$ as
$\psi(\vec{q},t) = \sum_{l=0}^{\infty} u_l(r,t) P_l(\cos \theta)/r$.
One can then 
modify the Schr\"odinger equation for $u_l(r,t)$ in the same way as Eq.$\,$(\ref{sch_1dkostin}) to incorporate a frictional force,
\begin{equation}
i\hbar \frac{\partial}{\partial t} u_l(r,t) =
\left[ - \frac{\hbar^2}{2m} \frac{\partial^2}{\partial r^2} + \frac{\hbar^2}{2m} \frac{l(l+1)}{r^2} + V(r) + \int^r dr_1 \, \gamma(r_1) \frac{\partial}{\partial r_1} S_l(r_1,t) \right]
u_l(r,t),
\label{sch_3dkostin}
\end{equation}
where $S_l(r,t)$ is the phase of the radial wave function,
$u_l(r,t) = |u_l(r,t)| \exp(i S_l(r,t) / \hbar)$.
We have here assumed a spherically symmetric potential, $V(\vec{q}) = V(r)$.
Notice that only the radial friction is taken into account here, 
while the angular momentum dissipation is neglected.
In the following, we only consider the s-wave scattering, $l=0$.

In applying Eq. (\ref{sch_3dkostin}) to scattering problems,
one needs to use the time-dependent approach,
since the Hamiltonian depends explicitly on time.
To this end, we propagate a wave packet and observe
how it bifurcates after it crosses the potential region.
Since a wave packet is a superposition of various energy waves,
one has to choose carefully the initial condition
to get scattering quantities at certain initial energy.
Notice that the energy projection approach \cite{Yabana97} is
inapplicable for our purpose, since the energy is not conserved.

In the initial condition, the width of the energy distribution
must be small enough to get reasonable results.
In this context, the energy means the expectation value
of the asymptotic Hamiltonian, $H_0$.
If $V(r)$ rapidly vanishes as $r \to \infty$, one can simply 
take the kinetic energy operator as $H_0$, that is, 
$H_0 = - \hbar^2/2m \left( \partial^2/\partial r^2 \right)$.
The minimum uncertainty wave packet in this case
has been discussed in Ref. \cite{Bracher11}, which reads
\begin{equation}
u_0^{\rm min}(r,t=0) \propto r e^{-(r-r_0)^2/4\sigma_r^2} e^{ik_0r},
\label{u_minimum}
\end{equation}
where $r_0$ and $\sigma_r$ are related to the mean position
and the width of the wave function in the coordinate space, respectively,
and $k_0$ is related to the mean initial energy.

In nuclear collisions, on the other hand, the potential is
a sum of the nuclear potential $V_N$ and the Coulomb potential
$V_C(r) = Z_PZ_Te^2/r$ with the projectile charge $Z_P$ and
the target charge $Z_T$. 
Since the Coulomb potential is a long range potential,
the asymptotic Hamiltonian $H_0$ has to include it, that is, 
$H_0 = - \hbar^2/2m \left( \partial^2/\partial r^2 \right) + V_C$.
Thus, the minimum uncertainty wave packet in the form of 
Eq. (\ref{u_minimum}) would not
be efficient in this case.
Instead, one needs to construct a wave packet from
the energy distribution, $f_{C}(E)$, 
of $H_0$ in the presence of the Coulomb potential. 
In analogy to the spherical Bessel functions, we find that this can be achieved as 
\begin{equation}
u_0^{C}(r,t=0) \propto \int_0^{\infty} dk \,
F_0(\eta, kr) e^{ikr_0} \sqrt{k f_{C}(E)}
\label{u_coulomb}
\end{equation}
with $E=\hbar^2k^2/2m$, 
where $\eta = mZ_PZ_Te^2/\hbar^2k$ is the Sommerfeld parameter and $F_0(\eta, kr)$ 
is the regular Coulomb wave function. 

With the initial condition given by Eq. (\ref{u_coulomb}), we compute the
penetrability of the Coulomb barrier 
for the $^{16}$O +  $^{208}$Pb system 
in the presence of friction. 
For the nuclear potential, we employ the optical potential in 
Ref. \cite{Eve08}, that is, 
\begin{equation}
V_N(r) = \frac{V_0}{1+\exp((r-R_{\rm v})/a_{\rm v})} +
i\frac{W_0}{1+\exp((r-R_{\rm w})/a_{\rm w})},
\label{optical_pot}
\end{equation}
with $V_0 = -901.4$ MeV, $R_{\rm v} = 8.44$ fm,
$a_{\rm v} = 0.664$ fm, $W_0 = -30$ MeV,
$R_{\rm w} = 6.76$ fm, and $a_{\rm w} = 0.4$ fm.
With this potential, the Coulomb barrier height $V_B$ is found to be 
$74.5$ MeV.

For a friction coefficient $\gamma(r)$, we employ the surface friction model \cite{FL96},
\begin{equation}
\gamma(r) = \frac{\gamma_0}{m} \left( V_B \frac{df_{\rm WS}}{dr} \right)^2,
\end{equation}
with the Woods-Saxon from factor $f_{\rm WS} = 1/(1+\exp((r-R_{\rm v})/a_{\rm v}))$. 
This is a general form of the friction coefficient obtained perturbatively \cite{GK78},
and has successfully been applied to above barrier fusion reactions
and to deep inelastic scatterings \cite{FL96}.
We arbitrarily set $\gamma_0 = 4.7 \times 10^{-23}$ s/MeV, which 
was used in the classical calculations.
We compute the phase of the wave function in the same way as in Ref. \cite{TH17}.

For the initial energy distribution in Eq. (\ref{u_coulomb}), we assume
the Gaussian form,
\begin{equation}
f_{C}(E) = \frac{1}{\sqrt{2\pi\sigma_E^2}}\,e^{-(E-E_0)^2/2\sigma_E^2}.
\label{ed}
\end{equation}
where $E_0$ and $\sigma_E$ are the mean and the width of
the initial energy distribution.
We have confirmed that $\sigma_E = 0.5$ MeV is sufficient
in the present parameter set.

\begin{figure}[bt]
\begin{center}
\includegraphics[width=7cm]{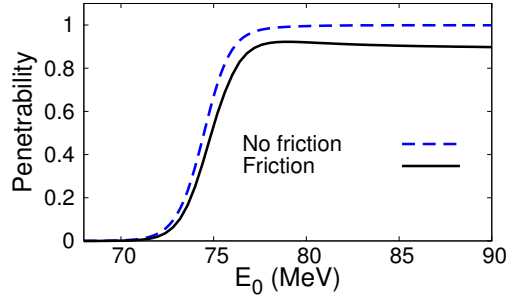} 
\end{center}
\caption{ 
Energy dependence of the penetrability of the Coulomb barrier for the $^{16}$O+$^{208}$Pb reaction. 
The dashed line shows the result without friction, while the solid line is for the result with friction.
}
\label{fig:wptun}
\end{figure}

Fig. \ref{fig:wptun} compares the penetrability obtained with and without friction.
One finds that the penetrability with friction is shifted to higher energies around the barrier.
That is, in the presence of friction, a particle needs an additional energy
to penetrate the barrier, which is originated from the energy dissipation.
One can also see that the penetrability does not reach unity at high energies,
but it is almost saturated at around $0.9$.
This means that the exit channel is in a quantum superposition
state of absorption and reflection even at sufficiently above barrier energies.
Notice that, in classical mechanics without a fluctuation force,
the penetrability can be only 0 or 1.
In this sense, this is peculiar to the quantum friction model.

In low-energy fusion reactions, it has been known
that low-lying collective excitations during the process
play a crucial role \cite{HT12}.
In the calculation shown in this paper, 
a part of such effect is implicitly taken into account in the nuclear potential.
An explicit treatment of low-lying collective excitations
together with a frictional force has been carried out in Ref. \cite{NN},
in which the experimental fusion cross sections for the $^{16}$O+$^{208}$Pb system 
are well reproduced from subbarrier to above barrier energies.
In that calculation, the same behavior as in Fig. \ref{fig:wptun} has been found,
which may be a key to achieve a consistent description. 

\section{System-plus-bath model}

We next consider a more microscopic model for quantum friction, employing 
a system-plus-bath model. 
To be more specific, we consider the Caldeira-Leggett model \cite{CL83}, whose Hamiltonian is given by, 
\begin{eqnarray}
H_{\rm tot} &=& H_S + \sum_{i} \hbar\omega_ia_i^\dagger a_i + h(q) \sum_i d_i(a_i^\dagger + a_i), 
\label{Htot}
\\
&\equiv& H_S + H_B+V_{\rm coup}, 
\end{eqnarray}
where 
$H_S$ and $H_B$ are the Hamiltonians for the system and the bath degrees of freedom, respectively, 
while $V_{\rm coup}$ is the coupling Hamiltonian between the system and the bath. 
Here, the bath degree of freedom is assumed to be a set of harmonic oscillators, whose 
creation and annihilation operators are denoted by 
and $a_i^\dagger$ and $a_i$, respectively. 
The coupling Hamiltonian is assumed to be
separable between the system and the bath degrees of freedom. In there, 
$d_i$ is the coupling strength, and $h(q)$ is the coupling form 
factor, where $q$ is the coordinate of the system. 

There are several ways to solve the Calderira-Leggett Hamiltonian. 
In Ref. \cite{CL83}, the bath degrees of freedom are integrated out 
using the path integral in order to obtain an effective action for the 
system degree of freedom (see also Ref. \cite{TB84}). 
One can also introduce the 
influence functional \cite{BT85}. 
Here we discuss the coupled-channels approach \cite{TH19}. 

In the coupled-channels approach \cite{HT12}, one expands the total wave function in terms of 
the eigen-wave functions of $H_B$, that is, 
\begin{equation}
\Psi_{\rm tot}(q,t)=\sum_{\{n_i\}}\psi_{\{n_i\}}(q,t)\,|\{n_i\}\rangle,
\label{wf_tot0}
\end{equation}
where the basis states $|\{n_i\}\rangle$ are given by, 
\begin{equation}
|\{n_i\}\rangle = \prod_i\frac{1}{\sqrt{n_i!}}\,\left(a_i^\dagger\right)^{n_i}|0\rangle. 
\end{equation}
Here, $|0\rangle$ is the vacuum state defined as $a_i|0\rangle = 0$. 
One can derive the coupled equations for $\psi_{\{n_i\}}(q,t)$ by evaluating the equation, 
\begin{equation}
\langle\{n_i\}|i\hbar\frac{\partial}{\partial t}|\Psi_{\rm tot}\rangle = 
\langle\{n_i\}|H_{\rm tot}|\Psi_{\rm tot}\rangle,
\end{equation}
that is, 
\begin{equation}
i\hbar\frac{\partial}{\partial t}\psi_{\{n_i\}}(q,t)
=\left(H_S+\sum_in_i\hbar\omega_i\right)\psi_{\{n_i\}}(q,t)
+\sum_{\{n'_i\}}
\langle\{n_i\}|V_{\rm coup}|\{n'_i\}\rangle\,
\psi_{\{n'_i\}}(q,t). 
\label{cc0}
\end{equation}

The coupled-channels equations, Eq. (\ref{cc0}), can be numerically solved when the number of the 
oscillator modes is not large \cite{HT12,HRK99}. However, in general, the number of the oscillator modes can be huge, or 
the distribution of the frequency of the 
oscillator may even be given as a continuous function. 
In that situation, it is almost hopeless to solve the coupled-channels equations directly. 
In order to overcome this problem, we introduce a more efficient basis to expand the total wave function \cite{TH19}.  
To this end, we first expand the function $e^{-i\omega t}$ with a finite basis set as, 
\begin{equation}
e^{-i\omega t}\sim\sum_{k=1}^K\eta_k(\omega)u_k(t), 
\label{expansion}
\end{equation}
where $u_k(t)$ is a known function such as a Bessel function, and $\eta_k(\omega)$ is the expansion coefficient. 
We then introduce a new phonon creation operator as, 
\begin{equation}
b_k^\dagger=\sum_i\frac{d_i}{\hbar}\eta_k(\omega_i)a_i^\dagger.
\end{equation}
Notice that the number of $k$ is finite, $k$ running from 1 to $K$, even though the number of $i$ may be infinite. 
We then construct the basis states using the operators $b_k^\dagger$, and expand the total wave function with 
them. That is, instead of Eq. (\ref{wf_tot0}), we expand the total wave function as, 
\begin{equation}
\Psi_{\rm tot}(q,t)=\sum_{\{\tilde{n}_k\}}\tilde{\psi}_{\{\tilde{n}_k\}}(q,t)\,
|\{\tilde{n}_k\}\rangle,
\label{Psi_tot}
\end{equation}
with 
\begin{equation}
|\{\tilde{n}_k\}\rangle = \prod_{k=1}^K
\frac{1}{\sqrt{\tilde{n}_k!}}\,\left(b_k^\dagger\right)^{\tilde{n}_k}|0\rangle. 
\end{equation}
One can then obtain the coupled-channels equations similar to Eq. (\ref{cc0}), that is, 
\begin{equation}
i\hbar\frac{\partial}{\partial t}\tilde{\psi}_{\{\tilde{n}_k\}}(q,t)
=H_S\,\tilde{\psi}_{\{\tilde{n}_k\}}(q,t)
+\sum_{\{\tilde{n}'_k\}}
\langle\{\tilde{n}_k\}|H_B+V_{\rm coup}|\{\tilde{n}'_k\}\rangle\,
\tilde{\psi}_{\{\tilde{n}'_k\}}(q,t). 
\label{cc}
\end{equation}
We once again emphasize that the dimension of the coupled-channels equations, Eq. (\ref{cc}), is much smaller 
than that of the original equations, Eq. (\ref{cc0}). 

The structure of the coupled-channels equations, Eq. (\ref{cc}), becomes simple when 
the basis functions $u_k(t)$ satisfy the following two conditions. 
\begin{enumerate}
\item The matrix $D$ defined as 
\begin{equation}
D_{kk'}\equiv \frac{1}{\hbar^2}\,\sum_id_i^2 \eta_k(\omega)\eta^*_{k'}(\omega),
\label{matrixD}
\end{equation}
is diagonal with respect $k$ and $k'$. 
That is, $D_{kk'}=\lambda_k\delta_{k,k'}$. 
Notice that the matrix $D$ can be expressed also as 
\begin{equation}
D_{kk'}\equiv \frac{1}{\hbar}\int^\infty_{-\infty}d\omega\,J(\omega) \eta_k(\omega)\eta^*_{k'}(\omega),
\end{equation}
with the spectral density given by, 
\begin{equation}
J(\omega)=\frac{1}{\hbar}\sum_id_i^2\delta(\omega-\omega_i).
\label{spectral-density}
\end{equation}

\item 
The basis function $u_k(t)$ is closed under differentiation, that is, 
\begin{equation}
\frac{du_k(t)}{dt}=\sum_{k'=1}^K\,C_{kk'}u_{k'}(t). 
\end{equation}
Notice that Bessel functions satisfy this condition since the following relation holds, 
\begin{equation}
\frac{d}{dx}J_k(x)=-\frac{1}{2}J_{k+1}(x)+\frac{1}{2}J_{k-1}(x),
\end{equation}
(with $J_{-k}(x)=(-1)^kJ_k(x)$ for an integer value of $k$). 

\end{enumerate}
See Eq. (31) in Ref. \cite{TH19} for the explicit form of the coupled-channels equations, 
Eq. (\ref{cc}). Ref. \cite{TH19} also provides an alternative derivation of the coupled-channels equations, which 
uses the influence functional of the path integral method. This allows one to extend the present 
formalism to finite temperatures.

\begin{figure}[bt]
\begin{center}
\includegraphics[width=6cm]{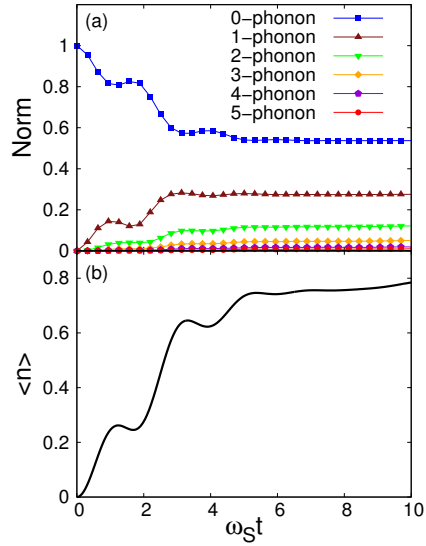}
\end{center}
\caption{ 
(The upper panel) The norm for each phonon number, $N=\sum_{k=1}^K\tilde{n}_k$. 
The solid line with squares, triangles, inverted triangles, diamonds, pentagons, and circles are for $N$ = 0, 1, 2, 3, 4, and 5, respectively. These are obtained by solving the 
coupled-channels equations 
with the phonon states up to $N_{\rm max} = 5$, for which 
the phonon operators are defined with the Bessel function basis with $K = 20$. 
(The Lower panel) The expectation value of the number of phonon. }
\label{fig:damped1}
\end{figure}

\begin{figure}[bt]
\begin{center}
\includegraphics[width=10cm]{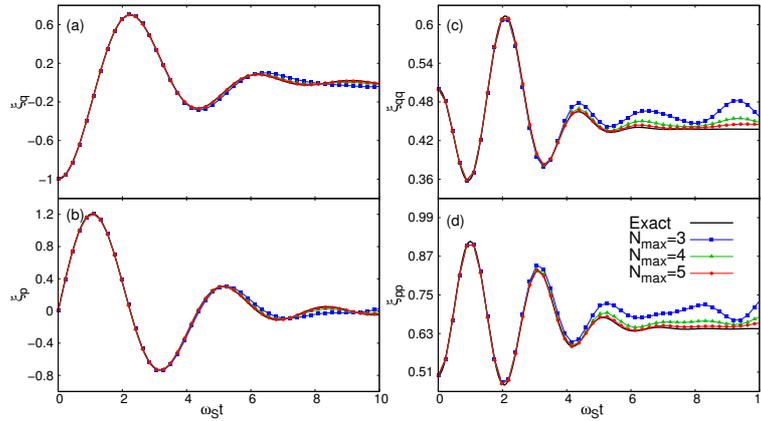}
\end{center}
\caption{ 
Comparison between the present method and the exact results for 
the expectation values of several quantities, that is, 
$\xi_q=\braket{q}/q_S$,  
$\xi_p=\braket{p} q_S /\hbar$, 
$\xi_{qq}=\braket{\left( q-\braket{q} \right)^2}/q_S^2$, 
and $\xi_{pp}=\braket{\left( p-\braket{p} \right)^2} q_S^2/\hbar^2$. 
The solid lines show the exact results, while the solid lines with squares, triangles, and circles are the results of the present 
method with $N_{\rm max} = 3$, $4$, and $5$, respectively. }
\label{fig:damped2}
\end{figure}

Figures \ref{fig:damped1} and \ref{fig:damped2} 
show results of a time evolution for a damped harmonic 
oscillator \cite{TH19}, 
for which we take the Hamiltonian for the system, $H_S$ in Eq. (\ref{Htot}), as 
\begin{equation}
\label{HS dho}
H_S = \frac{p^2}{2M} + \frac{1}{2} M \omega_S^2 q^2 + h^2(q) \sum_i \frac{d_i^2}{\hbar \omega_i},
\end{equation}
where $M$ and $\omega_S$ are the mass and the 
frequency of the system, respectively, and the last term represents 
the so called counter term.
In the following, we measure the length of the system in units of the 
the oscillator length $q_S$ defined by $q_S \equiv \sqrt{\hbar / M \omega_S}$, 
and the take the coupling form factor, $h(q)$, to be $h(q) = q/q_S$. 
We assume that the bath oscillators are distributed according to the 
spectral density (see Eq. (\ref{spectral-density})) as 
\begin{equation}
J(\omega) = V_I \,\frac{\omega}{\Omega} \,\sqrt{1-\left(\frac{\omega}{\Omega}\right)^2}. 
\end{equation}
In the numerical calculations shown below, we take 
$\hbar \omega_S = 2 \ {\rm eV}$, $V_I = 1 \ {\rm eV}$, 
and $\hbar \Omega = 4 \ {\rm eV}$. 

At $t=0$, we assume that 
$\tilde{\psi}_{\{\tilde{n}_k\}}(q,t=0)=0$ for $N\equiv\sum_{k=1}^K\tilde{n}_k\neq0$. 
For $N=0$, that is, for $\tilde{n}_k=0$ for all $k$, 
we assume that the wave function is given by, 
\begin{equation}
\tilde{\psi}_{N=0}(q,t=0)
=\frac{1}{\sqrt[4]{2 \pi \sigma_0^2}} \, e^{-(q-q_0)^2 / 4 \sigma_0^2} \, e^{i p_0 q / \hbar},
\end{equation}
with $q_0/q_S = -1$, $\sigma_0/q_S = 1/\sqrt{2}$, and $p_0 q_S / \hbar = 0$. 

The upper panel of Fig. \ref{fig:damped1} shows 
the norm for each phonon state $N$ as a function 
of $\omega_St$. Here, the norm is defined as, 
\begin{equation}
{\mathcal N}_N(t)\equiv
\sum_{\{\tilde{n}_k\}}\int dq\,|\tilde{\psi}_{\{\tilde{n}_k\}}(q,t)|^2 
\delta_{\sum_k\tilde{n}_k,N}.
\end{equation}
To draw this figure, we take Bessel functions, $J_k(\Omega t)$, 
for $u_k(t)$ 
in Eq. (\ref{expansion}) with $K=20$. A new basis is then constructed by 
diagonalizing the matrix $D$ in Eq. (\ref{matrixD}). 
With this basis, we solve the coupled-channels 
equations by including the phonon states with $N\leq 5$. 
The expectation value of the norm is also shown in the lower panel. 
As is expected, the number of phnon in the bath gradually increases as 
a function of time. 
Notice that 
the contribution of the 5-phonon states is small in the whole time range 
shown in the figure.
This justifies the truncation at $N_{\rm max} = 5$ 
for the present parameter set. 
One can also see that the contribution of each phonon reaches its 
equilibrium at around $\omega_S t = 6$.

Fig. \ref{fig:damped2} compares the results of the present method with the exact 
solution for the quantum damped harmonic oscillator. 
To this end, 
we evaluate the expectation values for the following four quantities: 
$\xi_q \equiv \braket{q}/q_S$, $\xi_p \equiv \braket{p} q_S /\hbar$, 
$\xi_{qq} \equiv \braket{\left( q-\braket{q} \right)^2}/q_S^2$, 
and $\xi_{pp} \equiv \braket{\left( p-\braket{p} \right)^2} q_S^2/\hbar^2$.
We carry out the calculations with three different values of $N_{\rm max}$, 
that is,  $N_{\rm max}$ = 3, 4, and 5, and compare them with the 
exact results shown by the solid lines. 
One can see that all of the calculations with 
$N_{\rm max} = 3$, $4$, and $5$ reproduce the exact results 
up to $\omega_S t \sim 5$, for which $J_{20}(\Omega t)$ is negligibly small and thus 
the expansion in Eq. (\ref{expansion}) with Bessel functions up to $K=20$ (that is, up to 
$J_{19}(\Omega t)$) is reasonable. 
The deviation from the exact results become significant for larger values of $\omega_St$, 
especially for the second order moments, $\xi_{qq}$ and $\xi_{pp}$. 
This is a natural consequence of the fact that the larger number of phonon states are required 
to describe the finer structures. 

\section{Summary}

We have discussed two time-dependent methods for quantum friction. The first 
method is based on an effective Hamiltonian, which is constructed so that expectation values 
of operators obey a classical equation of motion with friction. Such Hamiltonian is in general 
time-dependent, and we have solved it with a time-dependent wave packet method. 
The other method is to start with a total Hamiltonian with both the system and the environmental 
degrees of freedom and then eliminate the environmental degree of freedom to derive an equation which 
the system degree of freedom obeys. For this approach, we have presented a new efficient basis 
for coupled-channels equations. 
These two methods are complimentary to each other. In the first method, whereas several parameters 
have to be determined phenomenologically, a required computational time is much shorter than the 
second method. On the other hand, the second method is based on a more microscopic Hamiltonian, 
and thus less empirical inputs are required, 
even 
though a computational time may be large. 
By combining these two approaches appropriately, one may be able to achieve a quantum description of 
heavy-ion deep inelastic collisions as well as fusion reactions to synthesize superheavy elements. 

\section*{Author Contributions}
The contribution of each of the authors has been
significant and the manuscript is the result of an even effort of both the authors. 

\section*{Funding}
This work was supported by Tohoku University Graduate Program on Physics for
the Universe (GP-PU), and JSPS KAKENHI Grant Numbers JP18J20565 and 19K03861.

\section*{Acknowledgments}
We would like to thank Denis Lacroix and Guillaume Hupin for useful discussions.

\end{document}